\documentclass[conference, a4paper]{IEEEtran}
\IEEEoverridecommandlockouts
\usepackage{cite}
\usepackage{amsmath,amssymb,amsfonts}
\usepackage{algorithmic}
\usepackage{graphicx}
\usepackage{textcomp}
\usepackage{xcolor}
\def\BibTeX{{\rm B\kern-.05em{\sc i\kern-.025em b}\kern-.08em
    T\kern-.1667em\lower.7ex\hbox{E}\kern-.125emX}}

\makeatletter
\def\endthebibliography{%
  \def\@noitemerr{\@latex@warning{Empty `thebibliography' environment}}%
  \endlist
}
\makeatother

\begin{document}
\bstctlcite{IEEEexample:BSTcontrol}

\title{NEUROPULS: NEUROmorphic energy-efficient secure accelerators based on Phase change materials aUgmented siLicon photonicS
\thanks{This project has received funding from the European Union’s Horizon Europe research and innovation programme under grant agreement No. 101070238. Views and opinions expressed are however those of the author(s) only and do not necessarily reflect those of the European Union. Neither the European Union nor the granting authority can be held responsible for them.}
}
\author{
\IEEEauthorblockN{Fabio~Pavanello$^1$, Cedric~Marchand$^1$, Ian~O'Connor$^1$, R\'egis~Orobtchouk$^1$, Fabien~Mandorlo$^1$, Xavier~Letartre$^1$,\\
Sebastien~Cueff$^1$, Elena~Ioana~Vatajelu$^2$, Giorgio~Di~Natale$^2$,Benot~Cluzel$^3$, Aurelien Coillet$^3$, Benoit~Charbonnier$^4$,\\
Pierre~No\'e$^4$, Frantisek~Kavan$^5$, Martin~Zoldak$^5$, Michal~Szaj$^5$, Peter~Bienstman$^6$, Thomas~{Van~Vaerenbergh}$^7$,\\
Ulrich~Ruhrmair$^8$, Paulo~Flores$^9$, Luis~Guerra~e~Silva$^9$, Ricardo~Chaves$^9$, Luis-Miguel~Silveira$^9$,
Mariano~Ceccato$^{10}$, \\  Dimitris~Gizopoulos$^{11}$, George~Papadimitriou$^{11}$, Vasileios~Karakostas$^{11}$, Axel~Brando$^{12}$, Francisco~J.~Cazorla$^{12}$, \\ Ramon~Canal$^{12,13}$,
Pau~Closas$^{14}$, Adri\'a Gusi-Amig\'o$^{14}$, Paolo~Crovetti$^{15}$,
Alessio~Carpegna$^{16}$, \\ Tzamn~Melendez~Carmona$^{16}$, Stefano~Di~Carlo$^{16}$, Alessandro~Savino$^{16}$}\\
\IEEEauthorblockA{
$^1$Univ. Lyon, Ecole Centrale de Lyon, INSA Lyon, Université Claude Bernard Lyon 1, CPE Lyon, CNRS, INL\\
$^2$Univ. Grenoble Alpes, CNRS, Grenoble INP, TIMA, 38000 Grenoble, France\\
$^3$ICB UMR CNRS 6303, Université de Bourgogne Franche-Comté, Dijon, France\\
$^4$Univ. Grenoble Alpes, CEA, LETI, Grenoble, France,
$^5$ARGOTECH, Náchod, Czech Republic\\
$^6$Ghent University - imec, Gent, Belgium, 
$^7$Hewlett Packard Labs, HPE Belgium, B-1831 Diegem, Belgium
\\
$^8$Physics Dept. LMU Munchen, Munchen, Germany,
$^9$INESC-ID Lisboa, Lisbon, Portugal\\
$^{10}$Department of Computer Science, University of Verona – Verona, Italy\\
$^{11}$Department of Informatics and Telecommunications, National and Kapodistrian University of Athens, Athens, Greece\\
$^{12}$Barcelona Supercomputing Center, Barcelona, Spain,
$^{13}$Universitat Politècnica de Catalunya, Barcelona, Spain\\
$^{14}$Albora Technologies SL, Barcelona, Spain\\ % had to change to Spain as per the proposal you're located in Spain with the subsidiary
$^{15}$Politecnico di Torino, Dept. of Electronics and Telecommunications (DET), Italy\\
$^{16}$Politecnico di Torino, Control and Computer Eng. Department, Italy
%Emails:  fabio.pavanello@cnrs.fr, alessandro.savino@polito.it
}
}
\maketitle
\begin{abstract}
This special session paper introduces the Horizon Europe NEUROPULS project, which targets the development of secure and energy-efficient RISC-V interfaced neuromorphic accelerators using augmented silicon photonics technology. 
Our approach aims to develop an augmented silicon photonics platform, an FPGA-powered RISC-V-connected computing platform, and a complete simulation platform to demonstrate the neuromorphic accelerator capabilities.
In particular, their main advantages and limitations will be addressed concerning the underpinning technology for each platform.
Then, we will discuss three targeted use cases for edge-computing applications: Global National Satellite System (GNSS) anti-jamming, autonomous driving, and anomaly detection in edge devices.
Finally, we will address the reliability and security aspects of the stand-alone accelerator implementation and the project use cases.
\end{abstract}
%\vspace{-0.5cm}
\section{Introduction}
% discussing the framework of the neuropuls project wrt motivations and call objective / Fabio
The increasing need to process large amounts of data has been a major driver for developing novel, energy-efficient computing architectures \cite{van_albada_performance_2018}.
Among the multitude of approaches to tackle energy efficiency requirements, brain-inspired neuromorphic architectures are one of the most viable solutions thanks to their inner absence of I/O bottleneck between memory and processing units and their more natural mapping to machine learning (ML) algorithms \cite{peng_neuromorphic_2018}.
Neuromorphic approaches are especially suitable for developing lightweight, low-latency, and low-power accelerators in line with the requirements sought by edge-computing systems for applications such as autonomous driving, Internet of Things (IoT) devices, and 5G networks.% \cite{Shuman2022}.
\begin{figure*}[t!]
    \centering
    \includegraphics[width=0.75\linewidth]{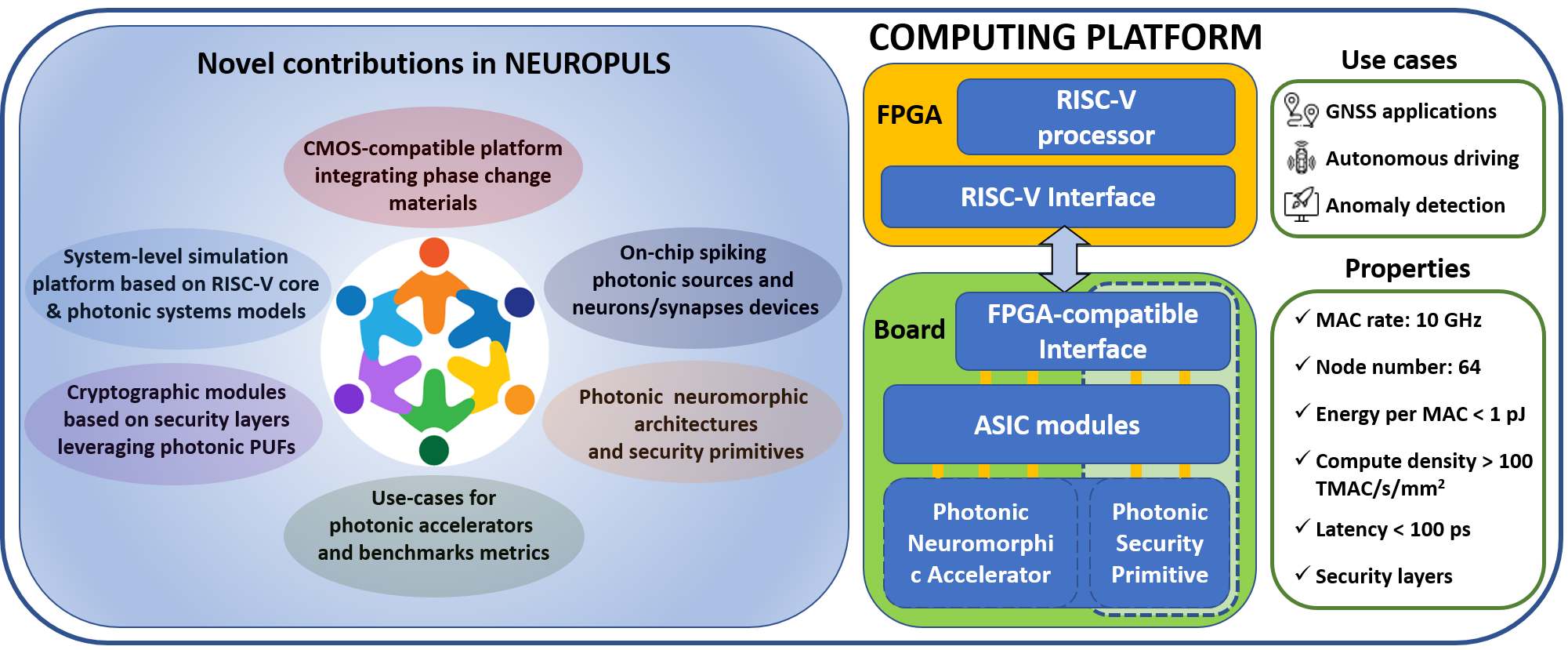}
    \caption{Overview of the novel contributions brought by the NEUROPULS project (left-side) and the computing platform that will be developed alongside its expected performance and targeted use cases (right-side).}
    \label{fig:concept3}
\end{figure*}

Integrating photonics can efficiently target all the needs above \cite{nahmias_photonic_2020} among the various technologies for building neuromorphic architectures.
This is mainly due to the current availability of mature CMOS-compatible platforms, e.g., Silicon-on-Insulator (SOI), which allow the dense integration of devices such as modulators, detectors, optical interfaces, etc., for volume scaling. 
In particular, photonic approaches can provide a key advantage concerning electronic ones regarding energy consumption, bandwidth, and latency thanks to low propagation losses in photonic integrated circuits (PICs) as well as the possibility for large parallelism using wavelength-multiplexing techniques, high-speed data encoding/decoding devices, and speed-of-light propagation throughout photonic components \cite{shen_deep_2017}.

In what follows, we will provide an overview of the NEUROPULS project and its scientific objectives towards strengthening the European digital supply chain and developing European secure specialized accelerator designs capable of delivering high-performance computing at ultra-low power operation. 
NEUROPULS’ unique technology will improve the performance per watt by at least two orders of magnitude for the targeted edge applications, thus addressing the performance reduction of current CMOS-based general-purpose computing platforms and the slow-down of Moore’s law.

%In this scenario, a wider range of photonic CMOS-compatible platforms will appear to address these different application domains using photonics, thus fostering cost reductions while establishing well-defined standards and certification guidelines, in our project for GNSS and autonomous driving use-cases.
%Lastly, novel and highly secure solutions based on hardware security layers at the silicon level leveraging photonic technologies will also be developed within the project.

\section{Proposed approach in NEUROPULS}
% 3 platforms very general - RISC-V etc. / Fabio
In NEUROPULS, we will develop three different platforms at technological, hardware computing, and simulation levels (see Fig.~\ref{fig:concept3}). The technological platform is based on the open-access silicon photonics platform from CEA-LETI with the addition of phase-change materials (PCMs) and III-V materials. 
The building blocks based on these technologies, such as plastic non-volatile photonic synapses \cite{rios_-memory_2019} and non-linear neurons based on spiking lasers \cite{ona_design_2021-1}, will enable various kinds of neuromorphic photonic architectures that NEUROPULS will investigate.

Security layers will be built upon Physical Unclonable Functions (PUFs), security primitives that avoid digital keys stored in memory \cite{pappu_physical_2002}. These primitives will be embedded in the photonic and electronic chips. A software/hardware interface will leverage the security layers at a silicon level to set up multiple security features for the accelerator, such as secure authentication, encryption of neural network (NN) data, and signing of the output data.

Using high-speed RF wiring and packaging, the photonic architectures will be driven by an ASIC chip linked to the PIC chip. The developed accelerator will then be interfaced by an FPGA where a RISC-V processor is implemented. The processor will be the interface between the outside world and the accelerator. It would allow transmission/retrieval of raw data/computing results and will handle the security protocols, leveraging the security primitives.
In our vision, the ASIC chip will also contain the RISC-V processor for large-volume manufacturing, leading to a stand-alone self-consistent hardware computing platform. However, for prototyping reasons, the ASIC in NEUROPULS will only contain the driving circuitry, the interface with the FPGA, and electronic PUFs. The latter will be used for hardware integrity to bundle the photonic and electronic chips uniquely. The PIC and the ASIC chips will be mounted on a board and connected using an FPGA interface to the FPGA hosting a RISC-V processor.

Computational models with different complexity levels for PCM-based photonic devices and systems will also be developed to simulate the behavior of the neuromorphic and PUF architectures taped out in the silicon photonics platforms and their response once included within the entire computing platform.
\begin{figure*}[t!]
    \centering
    \includegraphics[width=0.8\linewidth]{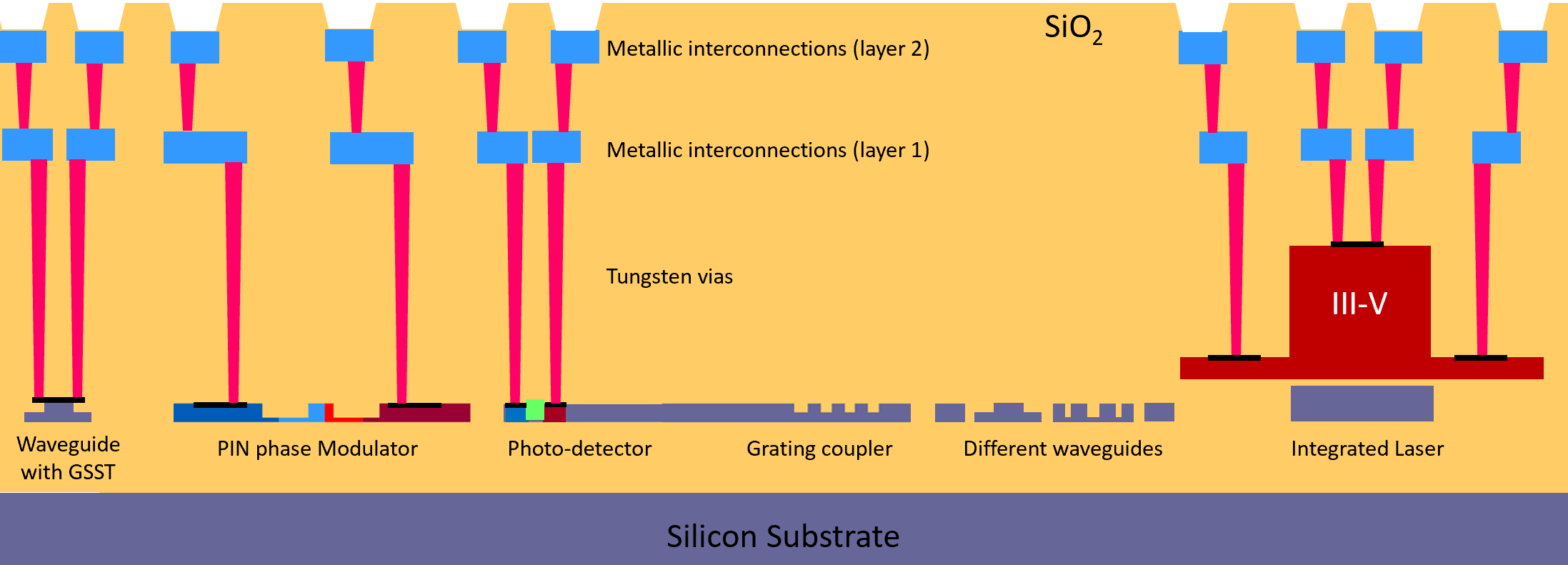}
    \caption{Cross-section of the augmented silicon photonics platform in NEUROPULS. PCMs such as GSST (left side) and III-V materials (right side) are integrated on top of waveguides. Two layers of interconnects will be available for routing purposes.}
    \label{fig:cross_section}
\end{figure*}
Finally, we will use the accelerator for three use cases of interest in edge computing: GNSS anti-jamming, autonomous driving, and anomaly detection in edge devices. State-of-the-art simulation tools considering the whole hardware computing platform will allow us to investigate performance scaling for the accelerator, e.g., energy consumption, latency, and speed as a function of the number of inputs/outputs/layers/neurons, etc. well its security features.
In NEUROPULS, we target highly demanding performance from our accelerator, which will allow us to improve by, e.g., two orders of magnitude the energy/MAC metric compared to concurring GPU technology for the selected edge-computing use cases. Besides, security features based on PUFs will be investigated for the stand-alone accelerator and the targeted use cases using both hardware and software approaches, from the security primitives level to the software-driven security protocols implemented. 

\section{CMOS-compatible platform} %0.5page
% technology oriented Benoit Charbonnier/Cluzel/Pierre/Seb/Xavier/Regis/Fabio
The platform that will be developed in the project is based on the silicon photonics platform from CEA-LETI, available in open-access multi-project wafer runs. It already includes several key building blocks, such as optical interfaces (grating couplers), modulators, detectors, resonators, etc. (see Fig.~\ref{fig:cross_section}). However, one of the novel features we aim to integrate includes PCMs. These thin PCM layers, e.g., GeSbTe (GST) and GeSSbTe (GSST) materials with different stoichiometric compositions, will be placed as thin (10-20-nm-thick) rectangular patches above optical waveguides as shown in Fig.~\ref{fig:cross_section}.
Their integration influences the optical field propagation due to changes in effective refractive index depending on the PCM state, i.e., amorphous versus crystalline. Strong modulation of the optical properties can be obtained due to the optical field interaction with the PCM patch. 
Various waveguides will be available, namely strip (fully etched) and rib (partially etched) configurations, where the overall silicon layer thickness is 300 nm. 
Two layers of metal interconnect will also be present to ease the routing of the components, and heaters will also be included (not present in the cross-section) to modify the crystalline degree of the PCM. Thanks to high-performance traveling-wave modulators and Ge detectors, the platform allows operating at frequencies above 50 GHz.

\section{Hardware computing platform} %0.5pages
% architectures and models Thomas/Peter/Fabio
Regarding the optical architecture, we will implement different options based on combining silicon photonics and PCMs.

One straightforward option is to consider an accelerator for matrix-vector products in feedforward NNs (FFNNs), where a mesh of modulators is programmed so that the system implements a certain matrix multiplication \cite{shen_deep_2017}. This can be done, e.g., by using singular value decomposition to factor the matrix into the product of a unitary matrix, a diagonal matrix, and another unitary matrix. 

The main limitation here is that the optical components' size prevents the implementation of huge matrices. Therefore, we will also explore other options for implementing larger systems. One approach we will consider is pruning, which involves removing unnecessary connections and weights in the network to reduce the overall size of the matrix. Another option is to use block matrix decompositions, where the matrix is divided into smaller blocks that can be processed separately. This allows for larger matrices to be implemented with a smaller number of optical components.

Furthermore, we will explore tensor-train approaches proposed in \cite{xiaoLargescaleEnergyefficientTensorized2021}. These approaches involve representing the matrix as a product of low-rank tensors, which can be more efficiently processed using optical components.
Using these various approaches, we aim to develop a scalable optical architecture that can handle larger matrices and more complex NNs. This will enable faster, more scalable, and more efficient training of NNs using photonics.
Apart from feedforward architectures, we will also study recurrent NNs (RNNs). We can consider fully trainable RNNs, but a variant called reservoir computing will also be considered. In reservoir computing, an RNN is randomly initialized and left untrained. Instead, what is trained is a linear combination of the time traces of the signals at each node. This technique has the advantage of being more computationally efficient than fully trainable RNNs, while still achieving good performance in many applications. This approach has been implemented in integrated photonics before \cite{vandoorne_experimental_2014}. It has been used, e.g., to realize nonlinear dispersion compensation of telecom signals \cite{Sackesyn:21}, demonstrating the potential of photonics in a wide range of applications beyond traditional ML.

We plan to incorporate PCMs in our proposed architecture to implement non-volatile optical weights. Various studies have used these materials, such as \cite{feldmann_all-optical_2019} and \cite{miscuglio_artificial_nodate}. Using non-volatile weights significantly reduces power consumption compared to volatile weights, which must be driven continuously or refreshed periodically.
However, in addition to their non-volatility, PCMs have another advantage we plan to exploit in our architecture: their nonlinear dynamics. E.g., by exciting the material with pulses rather than continuous-wave excitation, we can take advantage of the nonlinear behavior of the material to enable other computing paradigms, such as spiking NNs (SNNs). In such NNs, the neurons communicate using brief pulses or spikes rather than continuously varying signals. This enables the implementation of energy-efficient and highly parallel NNs.
To generate the spikes injected into the system, we will monolithically integrate lasers in III-V materials in the same platform to develop advanced high-extinction ratio (ER $>$ 8 dB) Q-switched spiking lasers \cite{10.3389/fphy.2022.1017714}. These hybrid III-V-on-Si spiking lasers are a scalable and low-cost alternative to previous pure III-V versions of these Q-switched lasers.

\section{Simulation platform} %1page
% architectures abstraction and modeling of accelerator behavior at a high system level -- scaling and performance prediction Alessandro-Stefano/DImitris/Miguel/Ian
Novel hardware architectures require the development of an ecosystem of state-of-the-art tools capable of supporting and promoting innovative hardware approaches. Specifically, the feasibility of innovation demands simulation tools that allow exploring the functional benefits of the new technology while supporting a precise estimation of the impact of the accelerators on the final system regarding performance, power consumption, and reliability, among others.

%NKUA
The NEUROPULS project aims to develop a simulation platform that models and evaluates computing systems incorporating neuromorphic accelerators and hardware security primitives. The simulation platform will explore the design space of heterogeneous computing systems that utilize photonic neuromorphic accelerators and hardware primitives to achieve optimal performance. The project will design and implement a toolchain for modeling and simulation at the system level to make photonics accelerators functional and programmable. The toolchain will automate the process of generating system-level models of photonic modules with varying levels of complexity and accuracy. The simulation platform will model a complete computing system, including multiple CPU cores, memory hierarchy, PNNs, and photonic security primitives (PUFs-based). The platform will also cover all computing stack layers, from hardware to application software. To evaluate the security provisions of photonics technology, the simulation platform will consider all software security aspects at the system and application software layers.

The platform will be built around gem5~\cite{9499847,10071105}: a state-of-the-art microarchitecture-level simulator widely used in many studies~\cite{9499847,10071105,9616430}. It will be enhanced with recent extensions supporting RISC-V-based systems modeling with a rich set of accelerators and flexible interfaces. The platform's flexibility will ease the exploration of trade-offs between security levels and corresponding performance, power consumption, and reliability penalties.

NEUROPULS will also investigate the reliability and power estimation of photonic hardware components. The NEUROPULS simulation platform aims to provide a flexible infrastructure that enables the comparison of various photonic accelerator parameters to determine neuromorphic hardware's power reduction and performance improvements. In addition, the platform will focus on RISC-V CPU cores and customized neuromorphic accelerators, along with their corresponding software stack, to evaluate the reliability of the complete heterogeneous system and identify potential reliability issues that different NN accelerators may face.
Therefore, NEUROPULS will create a simulation platform that is rapid, adaptable, and modular. It will be constructed using the gem5 simulator as its foundation and will support the evaluation of advanced NN accelerators and PUFs. 
%The simulation platform is expected to possess the following advanced characteristics:

%\begin{enumerate}
%    \item Allow any system designer to properly define the performance and power budget, as well as the security characteristics, for integrating the silicon photonics accelerators into a genuine system.
%    \item Develop accurate models of silicon photonics, NN accelerators, and photonic security hardware, using precise models driven by the neuromorphic hardware, for performance, area, reliability, and power analysis.
%    \item Design a flexible infrastructure and modular design that enables precise tuning of critical parameters and exploration of the design space, targeting different ISAs and microarchitectures (such as RISC-V or Arm).
%\end{enumerate}

\section{Targeted use-cases} %0.5page
% use-cases of relevance 

\subsection{GNSS anti-jamming}
% Pau/Adria
Satellite-based navigation systems, called GNSS, provide an accurate and reliable solution to position, velocity, and time (PNT) estimation in a myriad of applications \cite{morton2021position}. However, GNSS can be compromised by L-band intentional/unintentional jamming interferences, which typically overpower a GNSS receiver intending to deny its operation \cite{amin2016vulnerabilities}. 
Jamming episodes are not rare nowadays, making jammers cheap to buy and/or easy to build \cite{Ferre2020}, which threatens safety-critical applications or critical infrastructures relying on GNSS PNT.
Therefore, interference management is crucial to resilient GNSS usage, including detecting and classifying jamming occurrences. In this context, it has been shown in \cite{Ferre2019,lee_jamming_2020} that ML models provide promising results, including setups where those models are learned in a distributed federated learning manner \cite{WuPLANS23}. 
NEUROPULS targets this use case by enabling efficient NN-based classifiers to warn users of the presence/type of jamming interference.

\subsection{Autonomous driving}
%BSC
Trajectory prediction is fundamental to autonomous driving frameworks such as Apollo~\cite{apollo_modules} and Autoware~\cite{autoware}. Some trajectory prediction systems build to different extents on RNNs. For instance, Apollo uses some such networks as part of its trajectory prediction system. Other implementations building on Long Short-Term Memory (LSTM) are realized using multi-input multi-output LSTM-based RNNs~\cite{LSTMtrajectory}. Therefore, efficient RNN implementations bring new opportunities for trajectory prediction in autonomous driving systems.
NEUROPULS will use its acceleration technology to explore the efficiency of RNN-based trajectory prediction systems. The aim is to start with the simpler RNNs part of Apollo, mainly to test the approach's viability, and then scale to more complex LSTM-based RNNs to test the limits of the approach in the context of trajectory prediction.

\subsection{Anomaly detection in edge devices}
As an example of in-network computing using the photonic neuromorphic accelerator developed in the NEUROPULS project, we will target security applications by investigating the potential of NNs emulated by the photonic hardware to perform anomaly detection in a resource-constrained edge environment. Having these ultrafast neuromorphic accelerators embedded at the Edge to flag anomalies would enable identifying the malicious events where more expensive post-processing would otherwise be required. Possible use cases could be the prevention of IoT botnet, DDoS attacks \cite{Meidan2018,doriguzzi-corin_lucid_2020}, or in-vehicle network security \cite{Yassine2020}. By nature, anomaly detection is event-driven. Consequently, it is potentially a good match with the temporal information processing enabled by SNNs \cite{Chen2018,Yassine2020}, one of the NN architectures investigated during the NEUROPULS project. The aim is first to predict the performance of a variety of promising anomaly detection algorithms when adapted to the constraints present in photonic neuromorphic hardware using the hardware models and simulation framework developed in the project, where we will target benchmark tasks that will allow us to compare simulated classification accuracy, predicted energy-efficiency and problem size scalability with the reported performance of existing state-of-the-art algorithms running on traditional digital hardware as well as alternative emerging neuromorphic accelerators  \cite{Yan_2021}. 
%Subsequently, after selecting the most promising anomaly detection task for our photonic hardware, corresponding data related to an industrially relevant security application will be collected to run additional benchmarking tests on the experimental hardware prototype for a problem size that matches the size of our hardware demonstrator.  

\section{Reliability and security aspects} %1page
% reliability of PCM devices/architectures and security primitives/architecture Ioana/Ricardo/Ulrich/Mariano/Cedric/Fabio
In NEUROPULS, we will develop security layers at the silicon level based on PUFs. More precisely, we will leverage the already available technology for the photonic accelerator to implement novel architectures for building CMOS-compatible photonic PUFs. 
Such implementations are expected to be more robust against ambient fluctuations, aging, side-channel attacks, and machine-learning modeling than PUFs based on electronic technologies. 
This is due to the unique properties of PICs, such as lower dependence on temperature fluctuations or EM interference.  Most of their components, such as waveguides, do not present aging issues nor signal leakages throughout propagation (light is confined in a tight area below dielectric layers), and a large number of degrees of freedom (phase, polarization, amplitude, mode number, etc.) is available to enhance system complexity \cite{pavanello_recent_2021}.

Our research will investigate both so-called Weak PUFs and Strong PUFs \cite{ruhrmair2014pufs}:  The Weak PUF approach will use photonic PUF designs with very few challenges per PUF.  A system-specific secret key or a system-specific (secret key, public key)-pair will be derived from the (noisy) responses of these Weak PUFs.  Key aspects here are the already mentioned long-term stability of the responses, the use of optimal techniques for key extraction and error correction, and the statistical independence of the PUF responses for neighboring or adjacent photonic Weak PUFs.  To verify the latter, statistical tools like the NIST suite will be employed \cite{rukhin2001statistical}.  
Another important aspect is the secure processing of the derived key within the accelerator and the surrounding electronic system and circuits.%, for example, the required cryptomodules.
In the Strong PUF approach, a more complex photonic PUF structure must and will be used.  As required by Strong PUFs, it shall have a very large number of challenge-response pairs (CRPs) and a highly complex challenge-response (or input-output) relation.  Non-linear optical effects inside the photonic PUF must be exploited in this context.  

We will systematically test their response for complexity to verify the sufficient complexity (and thus security) of our photonic Strong PUF designs.  Several existing Strong PUF tests will be applied, and various new tests will be developed simultaneously.  They assess aspects such as the higher-order non-linearities in the PUF-responses, challenge bit sensitivity, or pseudo-randomness as measured by the NIST suite \cite{rukhin2001statistical}.  
Another key strategy for evaluating the security and complexity of our photonic Strong PUFs is the application of ML algorithms \cite{ruhrmair2013puf}.  Given a relatively small set of PUF-CRPs, we will assess whether ML algorithms can be trained to predict other yet unknown CRPs.  Critical figures here are the number of CRPs used in the ML algorithm's training phase, the computational effort in the training phase, and the prediction accuracy of the trained ML model.  We will apply different ML strategies, including the latest NNs and related techniques. 
We stress that on the protocol and application side, the two primitives of  Weak PUFs and Strong PUFs shall be mixed and interleaved, combining their mutual strengths:  Strong PUFs will allow identification and authentication of messages with long-term or short-term digital keys in the system.  Weak PUFs can be employed to derive secret digital keys whenever needed, at least avoiding the long-term presence of digital secrets in the hardware \cite{ruhrmair2022secret}.     

Such a high level of security at the hardware level will be exploited to secure the software level, i.e., the application code that the RISC-V processor runs. This will be achieved by designing well-defined APIs at the hardware-software interface to let the software layer request security services from the hardware layer. These services include encryption, decryption, and digital signing tasks, based on secret cryptographic keys that never leak to the software and thus to a potential malicious analyst~\cite{ceccato2019understanding}. On top of this, strong software hardening solutions would be developed, such as program obfuscation~\cite{ceccato2015largestudy} and tamper-detection~\cite{viticchie2016reactive} to limit malicious reverse engineering attacks to the software components.

\section{Conclusions} %+ references 1 page
The NEUROPULS project will address several research areas across the supply chain, from material science and photonic technologies to computing architectures, security layers, and high-level simulations.
Three different platforms will be developed: a silicon photonics-powered chip, an FPGA-powered RISC-V-connected platform, and a complete simulation ecosystem to demonstrate the neuromorphic accelerator capabilities. The development of such platforms will be pivotal for the NEUROPULS project to showcase a low-power secure accelerator based on an augmented silicon photonics platform addressing three different use cases of interest for edge computing: GNSS anti-jamming, autonomous driving, and anomaly detection in edge devices. We will target two orders of magnitude lower power consumption concerning state-of-the-art technology for the selected use cases.
Novel security layers based on the developed CMOS-compatible photonic platform will be developed by tightly integrating security primitives, i.e., PUFs, and security software interfaces. Such integration will enable various security features, which we will apply to the accelerator as stand-alone and for specific use cases.
The modular approach we propose in NEUROPULS will inform us about the technology scaling potential without relying solely on hardware prototypes but by accurate system modeling using gem5-based simulators.
The project's outcomes will foster an augmented interest in this technology and spur novel research directions and investments at a European level in photonics, which has been indicated as a key enabling technology for Industry 4.0 and RISC-V processors technology.

\bibliographystyle{IEEEtran}
\bibliography{neuromorphic,security,photonic_devices_systems,gnss}

\begin{thebibliography}{10}
\providecommand{\url}[1]{#1}
\csname url@samestyle\endcsname
\providecommand{\newblock}{\relax}
\providecommand{\bibinfo}[2]{#2}
\providecommand{\BIBentrySTDinterwordspacing}{\spaceskip=0pt\relax}
\providecommand{\BIBentryALTinterwordstretchfactor}{4}
\providecommand{\BIBentryALTinterwordspacing}{\spaceskip=\fontdimen2\font plus
\BIBentryALTinterwordstretchfactor\fontdimen3\font minus
  \fontdimen4\font\relax}
\providecommand{\BIBforeignlanguage}[2]{{%
\expandafter\ifx\csname l@#1\endcsname\relax
\typeout{** WARNING: IEEEtran.bst: No hyphenation pattern has been}%
\typeout{** loaded for the language `#1'. Using the pattern for}%
\typeout{** the default language instead.}%
\else
\language=\csname l@#1\endcsname
\fi
#2}}
\providecommand{\BIBdecl}{\relax}
\BIBdecl

\bibitem{van_albada_performance_2018}
\BIBentryALTinterwordspacing
S.~J. van Albada \emph{et~al.}, ``\BIBforeignlanguage{en}{Performance
  {Comparison} of the {Digital} {Neuromorphic} {Hardware} {SpiNNaker} and the
  {Neural} {Network} {Simulation} {Software} {NEST} for a {Full}-{Scale}
  {Cortical} {Microcircuit} {Model}},'' \emph{\BIBforeignlanguage{en}{Front.
  Neurosci.}}, vol.~12, p. 291, May 2018. [Online]. Available:
  \url{https://www.frontiersin.org/article/10.3389/fnins.2018.00291/full}
\BIBentrySTDinterwordspacing

\bibitem{peng_neuromorphic_2018}
\BIBentryALTinterwordspacing
H.-T. Peng \emph{et~al.}, ``\BIBforeignlanguage{en}{Neuromorphic {Photonic}
  {Integrated} {Circuits}},'' \emph{\BIBforeignlanguage{en}{IEEE J. Select.
  Topics Quantum Electron.}}, vol.~24, no.~6, pp. 1--15, Nov. 2018. [Online].
  Available: \url{https://ieeexplore.ieee.org/document/8364605/}
\BIBentrySTDinterwordspacing

\bibitem{nahmias_photonic_2020}
\BIBentryALTinterwordspacing
M.~A. Nahmias \emph{et~al.}, ``\BIBforeignlanguage{en}{Photonic
  {Multiply}-{Accumulate} {Operations} for {Neural} {Networks}},''
  \emph{\BIBforeignlanguage{en}{IEEE J. Select. Topics Quantum Electron.}},
  vol.~26, no.~1, pp. 1--18, Jan. 2020. [Online]. Available:
  \url{https://ieeexplore.ieee.org/document/8844098/}
\BIBentrySTDinterwordspacing

\bibitem{shen_deep_2017}
\BIBentryALTinterwordspacing
Y.~Shen \emph{et~al.}, ``\BIBforeignlanguage{en}{Deep learning with coherent
  nanophotonic circuits},'' \emph{\BIBforeignlanguage{en}{Nature Photon}},
  vol.~11, no.~7, pp. 441--446, Jul. 2017. [Online]. Available:
  \url{http://www.nature.com/articles/nphoton.2017.93}
\BIBentrySTDinterwordspacing

\bibitem{rios_-memory_2019}
\BIBentryALTinterwordspacing
C.~R{\'\i}os \emph{et~al.}, ``\BIBforeignlanguage{en}{In-memory computing on a
  photonic platform},'' \emph{\BIBforeignlanguage{en}{Sci. Adv.}}, vol.~5,
  no.~2, p. eaau5759, Feb. 2019. [Online]. Available:
  \url{https://www.science.org/doi/10.1126/sciadv.aau5759}
\BIBentrySTDinterwordspacing

\bibitem{ona_design_2021-1}
K.~M. Ona, B.~Charbonnier, and K.~Hassan, ``Design of an {Integrated}
  {III}–{V} on silicon semiconductor laser for spiking neural networks,'' in
  \emph{2021 {IEEE} {International} {Interconnect} {Technology} {Conference}
  ({IITC})}, Jul. 2021, pp. 1--2, iSSN: 2380-6338.

\bibitem{pappu_physical_2002}
\BIBentryALTinterwordspacing
R.~Pappu \emph{et~al.}, ``\BIBforeignlanguage{en}{Physical {One}-{Way}
  {Functions}},'' \emph{\BIBforeignlanguage{en}{Science}}, vol. 297, no. 5589,
  pp. 2026--2030, Sep. 2002. [Online]. Available:
  \url{https://www.science.org/doi/10.1126/science.1074376}
\BIBentrySTDinterwordspacing

\bibitem{xiaoLargescaleEnergyefficientTensorized2021}
\BIBentryALTinterwordspacing
X.~Xiao \emph{et~al.}, ``Large-scale and energy-efficient tensorized optical
  neural networks on iii--v-on-silicon moscap platform,'' \emph{APL Photonics},
  vol.~6, no.~12, p. 126107, 2021. [Online]. Available:
  \url{https://doi.org/10.1063/5.0070913}
\BIBentrySTDinterwordspacing

\bibitem{vandoorne_experimental_2014}
\BIBentryALTinterwordspacing
K.~Vandoorne \emph{et~al.}, ``\BIBforeignlanguage{en}{Experimental
  demonstration of reservoir computing on a silicon photonics chip},''
  \emph{\BIBforeignlanguage{en}{Nat Commun}}, vol.~5, no.~1, p. 3541, Mar.
  2014, number: 1 Publisher: Nature Publishing Group. [Online]. Available:
  \url{https://www.nature.com/articles/ncomms4541}
\BIBentrySTDinterwordspacing

\bibitem{Sackesyn:21}
\BIBentryALTinterwordspacing
S.~Sackesyn \emph{et~al.}, ``Experimental realization of integrated photonic
  reservoir computing for nonlinear fiber distortion compensation,'' \emph{Opt.
  Express}, vol.~29, no.~20, pp. 30\,991--30\,997, Sep 2021. [Online].
  Available: \url{https://opg.optica.org/oe/abstract.cfm?URI=oe-29-20-30991}
\BIBentrySTDinterwordspacing

\bibitem{feldmann_all-optical_2019}
\BIBentryALTinterwordspacing
J.~Feldmann \emph{et~al.}, ``\BIBforeignlanguage{en}{All-optical spiking
  neurosynaptic networks with self-learning capabilities},''
  \emph{\BIBforeignlanguage{en}{Nature}}, vol. 569, no. 7755, pp. 208--214, May
  2019. [Online]. Available:
  \url{http://www.nature.com/articles/s41586-019-1157-8}
\BIBentrySTDinterwordspacing

\bibitem{miscuglio_artificial_nodate}
M.~Miscuglio \emph{et~al.}, ``Artificial synapse with mnemonic functionality
  using gsst-based photonic integrated memory,'' in \emph{2020 International
  Applied Computational Electromagnetics Society Symposium (ACES)}, 2020, pp.
  1--3.

\bibitem{10.3389/fphy.2022.1017714}
\BIBentryALTinterwordspacing
K.~Mekemeza-Ona, B.~Routier, and B.~Charbonnier, ``All optical q-switched laser
  based spiking neuron,'' \emph{Frontiers in Physics}, vol.~10, 2022. [Online].
  Available:
  \url{https://www.frontiersin.org/articles/10.3389/fphy.2022.1017714}
\BIBentrySTDinterwordspacing

\bibitem{9499847}
G.~Papadimitriou and D.~Gizopoulos, ``Demystifying the system vulnerability
  stack: Transient fault effects across the layers,'' in \emph{2021 ACM/IEEE
  48th Annual International Symposium on Computer Architecture (ISCA)}, 2021,
  pp. 902--915.

\bibitem{10071105}
------, ``{AVGI: Microarchitecture-Driven, Fast and Accurate Vulnerability
  Assessment},'' in \emph{2023 IEEE International Symposium on High-Performance
  Computer Architecture (HPCA)}, 2023, pp. 935--948.

\bibitem{9616430}
P.~R. Bodmann \emph{et~al.}, ``Soft error effects on arm microprocessors: Early
  estimations versus chip measurements,'' \emph{IEEE Transactions on
  Computers}, vol.~71, no.~10, pp. 2358--2369, 2022.

\bibitem{morton2021position}
Y.~J. Morton \emph{et~al.}, \emph{{Position, Navigation, and Timing
  Technologies in the 21st Century: Integrated Satellite Navigation, Sensor
  Systems, and Civil Applications}}.\hskip 1em plus 0.5em minus 0.4em\relax
  John Wiley \& Sons, 2021.

\bibitem{amin2016vulnerabilities}
M.~G. Amin \emph{et~al.}, ``Vulnerabilities, threats, and authentication in
  satellite-based navigation systems [scanning the issue],'' \emph{Proceedings
  of the IEEE}, vol. 104, no.~6, pp. 1169--1173, 2016.

\bibitem{Ferre2020}
R.~Morales-Ferre \emph{et~al.}, ``A survey on coping with intentional
  interference in satellite navigation for manned and unmanned aircraft,''
  \emph{IEEE Communications Surveys Tutorials}, vol.~22, no.~1, pp. 249--291,
  2020.

\bibitem{Ferre2019}
R.~M. Ferre, A.~D.~L. Fuente, and E.~S. Lohan, ``{Jammer classification in GNSS
  bands via machine learning algorithms},'' \emph{Sensors (Switzerland)},
  vol.~19, no.~22, pp. 5--7, 2019.

\bibitem{lee_jamming_2020}
\BIBentryALTinterwordspacing
G.-H. Lee, J.~Jo, and C.~H. Park, ``\BIBforeignlanguage{en}{Jamming
  {Prediction} for {Radar} {Signals} {Using} {Machine} {Learning} {Methods}},''
  \emph{\BIBforeignlanguage{en}{Security and Communication Networks}}, vol.
  2020, pp. 1--9, Jan. 2020. [Online]. Available:
  \url{https://www.hindawi.com/journals/scn/2020/2151570/}
\BIBentrySTDinterwordspacing

\bibitem{WuPLANS23}
P.~Wu \emph{et~al.}, ``{Jammer classification with Federated Learning},'' in
  \emph{Proc. of the IEEE/ION PLANS}, Monterey, CA, April 2023.

\bibitem{apollo_modules}
\BIBentryALTinterwordspacing
ApolloAuto. (2018) {Apollo 3.0 Software Architecture}. [Online]. Available:
  \url{https://github.com/ApolloAuto/apollo/blob/master/docs/specs/Apollo}
\BIBentrySTDinterwordspacing

\bibitem{autoware}
T.~A. Foundation, ``{Autoware. An open autonomous driving platform.}''
  \url{https://github.com/CPFL/Autoware/}, 2016.

\bibitem{LSTMtrajectory}
M.~Cajamarca, ``{A Keras multi-input multi-output LSTM-based RNN for object
  trajectory forecasting},''
  \url{https://github.com/MarlonCajamarca/Keras-LSTM-Trajectory-Prediction},
  2022.

\bibitem{Meidan2018}
Y.~Meidan \emph{et~al.}, ``{N-BaIoT-Network-based detection of IoT botnet
  attacks using deep autoencoders},'' \emph{IEEE Pervasive Computing}, vol.~17,
  no.~3, pp. 12--22, 2018.

\bibitem{doriguzzi-corin_lucid_2020}
R.~Doriguzzi-Corin \emph{et~al.}, ``\BIBforeignlanguage{en}{Lucid: {A}
  {Practical}, {Lightweight} {Deep} {Learning} {Solution} for {DDoS} {Attack}
  {Detection}},'' \emph{\BIBforeignlanguage{en}{IEEE TRANSACTIONS ON NETWORK
  AND SERVICE MANAGEMENT}}, vol.~17, no.~2, p.~14, 2020.

\bibitem{Yassine2020}
Y.~Jaoudi, C.~Yakopcic, and T.~Taha, ``Conversion of an unsupervised anomaly
  detection system to spiking neural network for car hacking identification,''
  in \emph{2020 11th International Green and Sustainable Computing Workshops
  (IGSC)}, 2020, pp. 1--4.

\bibitem{Chen2018}
Q.~Chen \emph{et~al.}, ``{AnRAD: A neuromorphic anomaly detection framework for
  massive concurrent data streams},'' \emph{IEEE Transactions on Neural
  Networks and Learning Systems}, vol.~29, no.~5, pp. 1622--1636, 2018.

\bibitem{Yan_2021}
\BIBentryALTinterwordspacing
Y.~Yan \emph{et~al.}, ``Comparing loihi with a spinnaker 2 prototype on
  low-latency keyword spotting and adaptive robotic control,''
  \emph{Neuromorphic Computing and Engineering}, vol.~1, no.~1, p. 014002, jul
  2021. [Online]. Available: \url{https://dx.doi.org/10.1088/2634-4386/abf150}
\BIBentrySTDinterwordspacing

\bibitem{pavanello_recent_2021}
\BIBentryALTinterwordspacing
F.~Pavanello \emph{et~al.}, ``\BIBforeignlanguage{en}{Recent {Advances} in
  {Photonic} {Physical} {Unclonable} {Functions}},'' in
  \emph{\BIBforeignlanguage{en}{2021 {IEEE} {European} {Test} {Symposium}
  ({ETS})}}.\hskip 1em plus 0.5em minus 0.4em\relax Bruges, Belgium: IEEE, May
  2021, pp. 1--10. [Online]. Available:
  \url{https://ieeexplore.ieee.org/document/9465434/}
\BIBentrySTDinterwordspacing

\bibitem{ruhrmair2014pufs}
U.~R{\"u}hrmair and D.~E. Holcomb, ``Pufs at a glance,'' in \emph{2014 Design,
  Automation \& Test in Europe Conference \& Exhibition (DATE)}.\hskip 1em plus
  0.5em minus 0.4em\relax IEEE, 2014, pp. 1--6.

\bibitem{rukhin2001statistical}
A.~Rukhin \emph{et~al.}, ``A statistical test suite for random and pseudorandom
  number generators for cryptographic applications,'' Booz-allen and hamilton
  inc mclean va, Tech. Rep., 2001.

\bibitem{ruhrmair2013puf}
U.~R{\"u}hrmair \emph{et~al.}, ``Puf modeling attacks on simulated and silicon
  data,'' \emph{IEEE transactions on information forensics and security},
  vol.~8, no.~11, pp. 1876--1891, 2013.

\bibitem{ruhrmair2022secret}
U.~R{\"u}hrmair, ``Secret-free security: A survey and tutorial,'' \emph{Journal
  of Cryptographic Engineering}, vol.~12, no.~4, pp. 387--412, 2022.

\bibitem{ceccato2019understanding}
\BIBentryALTinterwordspacing
M.~Ceccato \emph{et~al.}, ``Understanding the behaviour of hackers while
  performing attack tasks in a professional setting and in a public
  challenge,'' \emph{Empirical Software Engineering}, vol.~24, no.~1, pp.
  240--286, Feb 2019. [Online]. Available:
  \url{https://doi.org/10.1007/s10664-018-9625-6}
\BIBentrySTDinterwordspacing

\bibitem{ceccato2015largestudy}
\BIBentryALTinterwordspacing
------, ``\BIBforeignlanguage{English}{A large study on the effect of code
  obfuscation on the quality of java code},''
  \emph{\BIBforeignlanguage{English}{Empirical Software Engineering}}, vol.~20,
  no.~6, pp. 1486--1524, 2015. [Online]. Available:
  \url{http://dx.doi.org/10.1007/s10664-014-9321-0}
\BIBentrySTDinterwordspacing

\bibitem{viticchie2016reactive}
\BIBentryALTinterwordspacing
A.~Viticchi{\'e} \emph{et~al.}, ``Reactive attestation: Automatic detection and
  reaction to software tampering attacks,'' in \emph{Proceedings of the 2016
  ACM Workshop on Software PROtection}, ser. SPRO '16.\hskip 1em plus 0.5em
  minus 0.4em\relax New York, NY, USA: ACM, 2016, pp. 73--84. [Online].
  Available: \url{http://doi.acm.org/10.1145/2995306.2995315}
\BIBentrySTDinterwordspacing

\end{thebibliography}
\end{document}